\documentclass[dvipdfmx]{jpsj-suppl}
\usepackage{txfonts} 

\title{
  Theoretical photo-disintegration of $^{16}$O
}

\author{Masahiko \textsc{Katsuma}}

\inst{Advanced Mathematical Institute, Osaka City University, Osaka 558-8585, Japan}

\email{mkatsuma@gmail.com; mkatsuma@sci.osaka-cu.ac.jp}

\recdate{August 19, 2016}

\abst{
  The photodisintegration of $^{16}$O is predicted to be dominated by $E$2 excitation in the vicinity of the $\alpha$-particle threshold.
  The reaction rates of $^{12}$C($\alpha$,$\gamma$)$^{16}$O are expected to be determined from this reaction.
}

\kword{Photodisintegration, $^{12}$C($\alpha$,$\gamma$)$^{16}$O, Helium burning, Nuclear clustering}

\begin{document}
\maketitle

\section{Introduction}
  The $^{12}$C($\alpha$,$\gamma$)$^{16}$O reaction plays the crucial role in the nucleosynthesis of elements in a star.
  However, the reaction energy corresponding to the helium burning temperature ($E_{c.m.}= 0.3$ MeV) is too low to obtain the cross section at the present laboratories \cite{Rol88}.
  In general, the low-energy cross section is extrapolated from the available experimental data, or it is inferred from the indirect measurements.

  The photodisintegration of $^{16}$O, i.e. the inverse reaction, is expected to give the experimental data more accurately from the higher probability of the reaction events.
  In addition, the angular distribution of the emitted $\alpha$-particle is considered to be important to reveal the reaction mechanism.
  The strong coupling feature involves the nuclear reaction in the complicated process for the compound nucleus, and it gives the strong interference between two states, 1$^-_1$($E_x= 7.12$ MeV) and 1$^-_2$ ($E_x= 9.59$ MeV).
  This interference has been thought to describe the cross section of $^{12}$C($\alpha$,$\gamma$)$^{16}$O at $E_{c.m.}= 0.3$ MeV ($E$1 transition).
  In contrast, I predict that the $E$2 transition dominates the low-energy cross section \cite{Kat08}.
  This is caused by the 2$^+_1$ state ($E_x= 6.92$ MeV), which is well described by the $\alpha$+$^{12}$C cluster structure (\cite{Kat10}, and references therein).
  The $E$1 transition is negligible because of the weak coupling and iso-spin selection rule.
  The large $\alpha$-particle width comes from the $\alpha$-cluster state.

  In this contribution, I show the theoretical result of the photoelectric cross section of $^{16}$O($\gamma$,$\alpha$)$^{12}$C \cite{Kat14} in the vicinity of the $\alpha$-particle threshold for the future experiments.
  I use the potential model (PM) and the previous result of $^{12}$C($\alpha$,$\gamma$)$^{16}$O \cite{Kat08,Kat10,Kat12}.

\section{Potential model and Results}
  Before moving on to the result, let me recall PM, the reaction mechanism, and the important energies for the reaction rates, briefly.

  The photoelectric cross section is given from the capture cross section, $\sigma_{\gamma\alpha}= [k_\alpha^2/(2k_\gamma^2)]\,\sigma_{\alpha\gamma}$.
  In general, the wavenumber $k_\gamma$ of photon is smaller than $k_\alpha$ of the incident $\alpha$-particle.
  So, the cross section $\sigma_{\gamma\alpha}$ of the photodisintegration is expected to be larger than $\sigma_{\alpha\gamma}$ of $^{12}$C($\alpha$,$\gamma$)$^{16}$O.
  The capture cross section is given as $\sigma_{\alpha\gamma}\propto\left|\langle\phi_f|\tilde{e}M^E_\lambda|\phi_i\rangle\right|^2$, where $M^E_\lambda$ is the electric operator of $E\lambda$ transition.
  To generate the scattering state $\phi_i$, I use the potential describing elastic scattering \cite{Kat10,Pla87,Ing94}.
  The $\alpha$+$^{12}$C cluster structure in $^{16}$O is semi-classically defined by the evaluation of refractive scattering \cite{Bra97,Sat83}.
  To generate the ground state $\phi_f$ of $^{16}$O, the potential strength is adjusted to reproduce the $\alpha$-particle separation energy \cite{Sat83}.
  The effective charge $\tilde{e}$ is obtained from the $\chi^2$ optimization of the astrophysical $S$-factor data \cite{Ass06,Kun02,Sch12}.
  The $S$-factor is used, instead of the cross section, to compensate for the rapid energy variation below the barrier, $S=E_{c.m.}\exp(2\pi\eta)\sigma_{\alpha\gamma}$; $\eta=12e^2/\hbar v$; $v$ is the relative velocity.
  The photoelectric cross section is predicted from the available result of $^{12}$C($\alpha$,$\gamma$)$^{16}$O.

\begin{figure}[t]
  \begin{center}
    \begin{tabular}{cc}
      \includegraphics[width=0.44\linewidth]{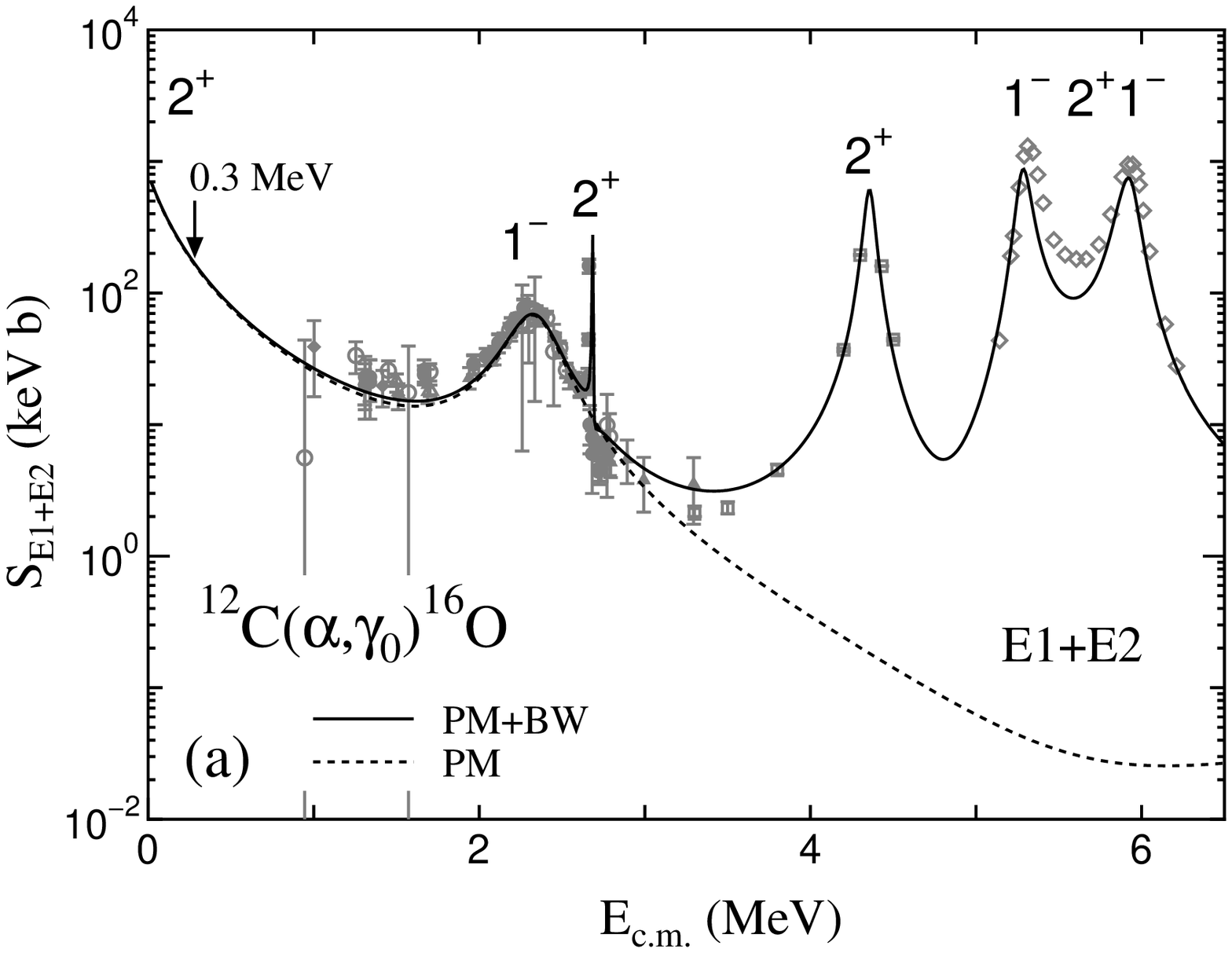} &
      \includegraphics[width=0.41\linewidth]{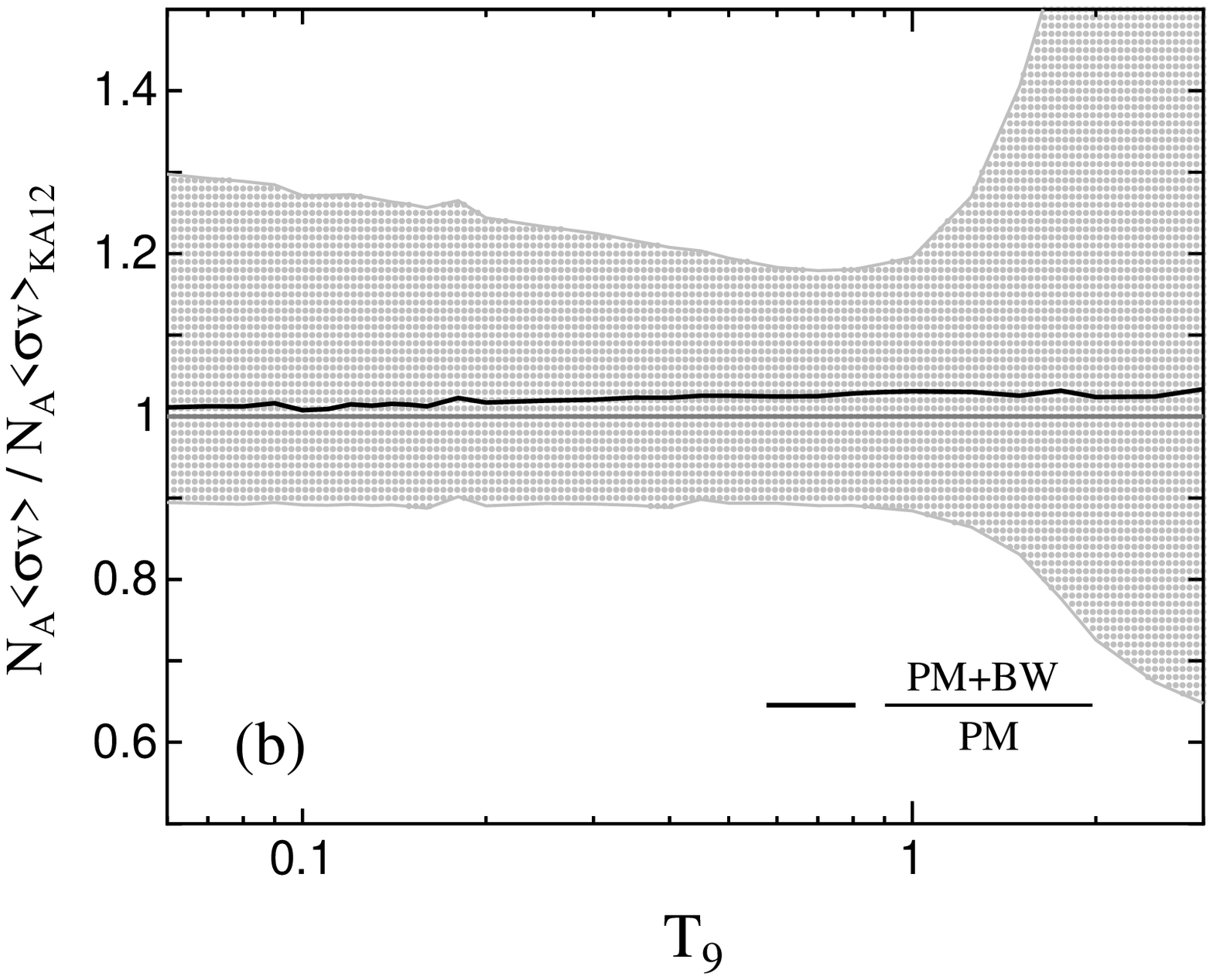}
    \end{tabular}
  \end{center}
  \caption{
    (a) Astrophysical $S$-factor of $E$1+$E$2 for $^{12}$C($\alpha$,$\gamma$)$^{16}$O.
    The solid curve is the result of PM with the additional experimental resonances in the Breit-Wigner form (PM+BW).
    The dotted curve is obtained from PM.
    The experimental data are taken from \cite{Ass06,Kun02,Sch12}.
    (b) The derived reaction rates of PM+BW in the ratio to PM.
    The shade aria is from the uncertainties of the parameter in PM \cite{Kat12}.
    The resonant contribution is negligible (less than 4\% difference), compared with the uncertainties below the barrier.
  }
  \label{fig:1}
\end{figure}

  I first review my previous result of the $S$-factor for $^{12}$C($\alpha$,$\gamma$)$^{16}$O, and articulate the reaction mechanism.
  The dotted curve in Fig.~\ref{fig:1}(a) is obtained from PM.
  To examine the reaction rates, the experimental resonances in the Breit-Wigner form are temporarily appended to PM. (PM+BW: solid curve)
  The arrow indicates the most important energy of helium burning temperature.
  The derived reaction rates from PM+BW are illustrated in the ratio to PM in Fig.~\ref{fig:1}(b).
  The contribution in the reaction rates from the additional resonances is negligible, although the large difference can be found above $E_{c.m.} = 3$ MeV in the $S$-factor.
  The reaction rates are thus determined from the direct capture component of PM below the barrier.
  In particular, the tail of the subthreshold 2$^+_1$ state ($E_{c.m.}=-0.245$ MeV) leads to the enhancement of the low-energy $S$-factor, which dominates the resultant reaction rates.
  The recent experimental results \cite{Ass06,Kun02} of the $\gamma$-ray angular distribution appear to advocate the weak coupling mechanism, and they seem to be made from the interference between two $\alpha$+$^{12}$C molecular states \cite{Kat08,Kat10}.
  The reaction rates are concordant with those from the strong coupling mechanism of the previous $R$-matrix analysis, because the total $S$-factors are comparable at $E_{c.m.}=0.3$ MeV \cite{Kat12}.

\begin{figure}[t]
  \begin{center}
    \begin{tabular}{cc}
      \includegraphics[width=0.48\linewidth]{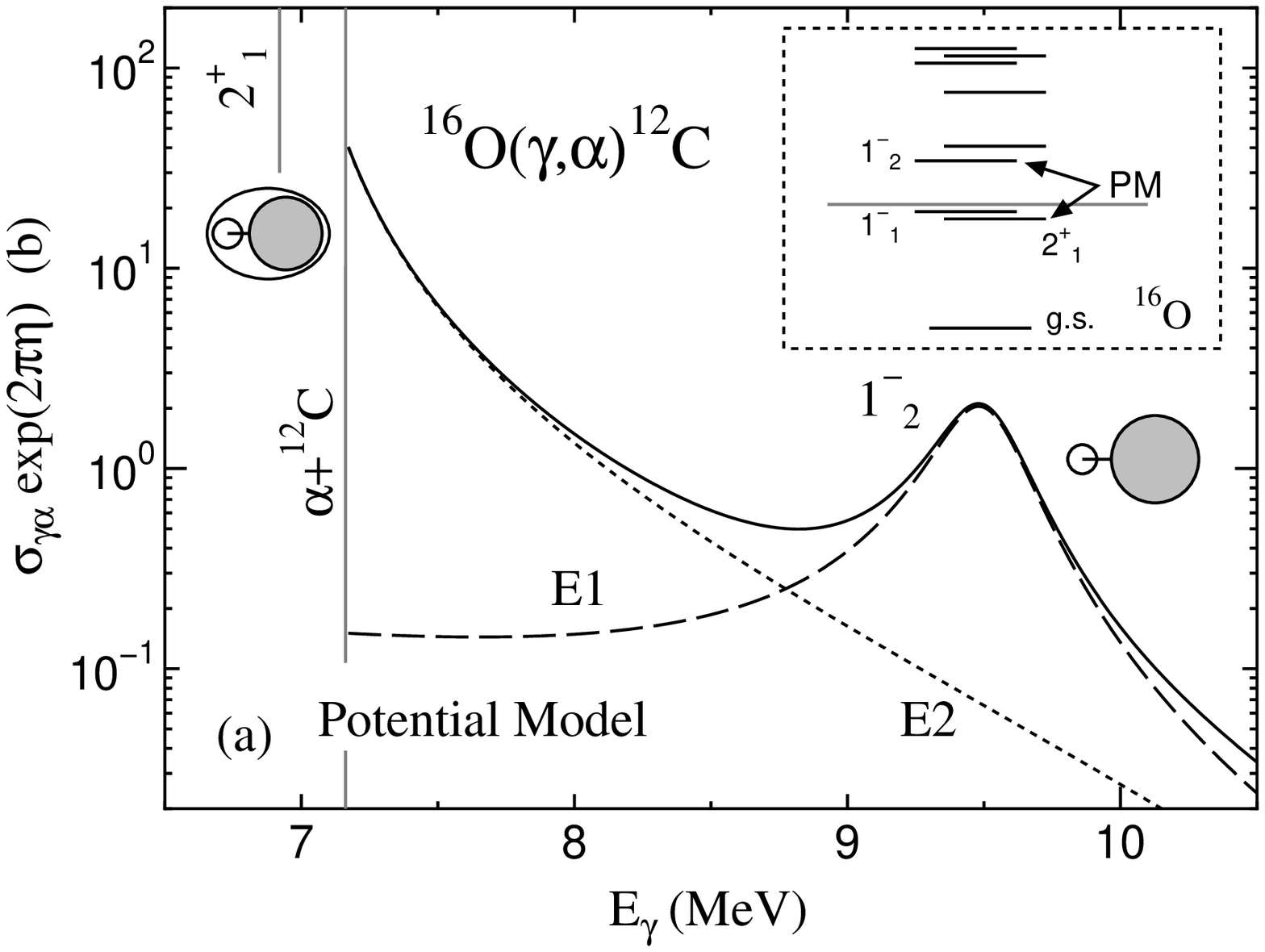} &
      \includegraphics[width=0.465\linewidth]{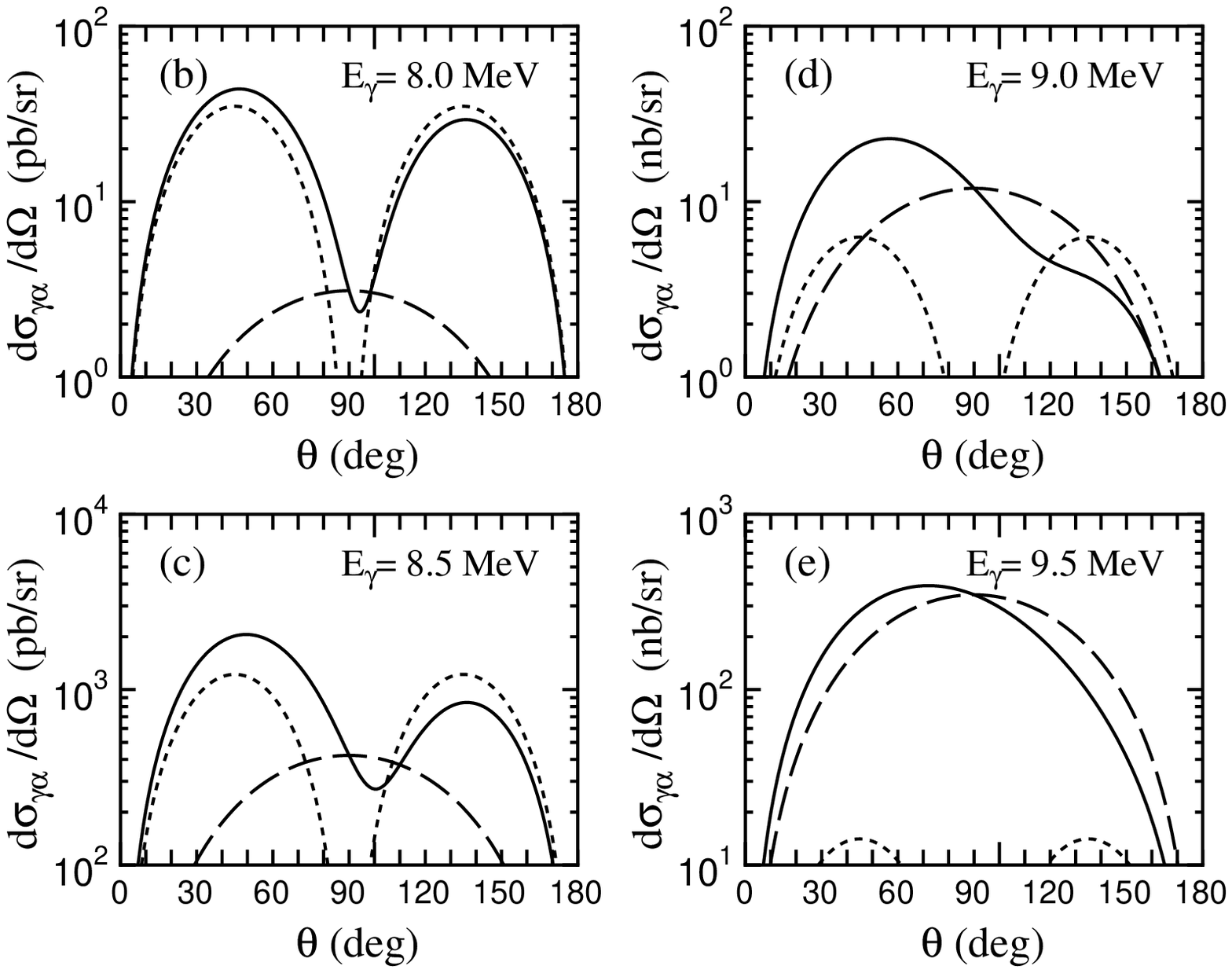} 
    \end{tabular}
  \end{center}
  \caption{
    Photoelectric cross section of $^{16}$O($\gamma$,$\alpha$)$^{12}$C \cite{Kat14}.
    (a) Integrated cross section, multiplied by a factor of $\exp(2\pi\eta)$.
    The $\alpha$-particle is emitted above the $\alpha$+$^{12}$C threshold ($E_\gamma=7.162$ MeV).
    The solid curve is the result of PM.
    The dashed and dotted curves are the $E$1 and $E$2 excitation, respectively.
    The angular distribution of the $\alpha$-particle at (b) $E_\gamma=8.0$, (c) 8.5, (d) 9.0, and (e) 9.5 MeV.
    The dashed and dotted curves represent the pure $E$1 and $E$2 components, respectively.
    The interference makes the asymmetric distribution. (solid curves)
  }
  \label{fig:2}
\end{figure}

  The photoelectric cross section is displayed in Fig.~\ref{fig:2}(a).
  The dashed and dotted curves are the $E$1 and $E$2 components, respectively.
  The solid curve is the sum of them.
  The cross section is multiplied by a factor of $\exp(2\pi\eta)$.
  From Fig.~\ref{fig:2}(a), the photodisintegration is found to be dominated by $E$2 excitation in the vicinity of $\alpha$-particle threshold.
  Figures~\ref{fig:2}(b)--\ref{fig:2}(e) show the angular distribution of the emitted $\alpha$-particle at $E_\gamma=8.0$, 8.5, 9.0, and 9.5 MeV.
  The solid curves are the results from PM.
  The dashed and dotted curves represent the pure $E$1 and $E$2 components.
  The single peak is made by $E$1 excitation to the 1$^-_2$ state at $E_x\approx 9.5$ MeV.
  At low $E_\gamma$, the angular distribution has the double peaks because of the 2$^+_1$ state.
  The interference between two molecular states, 2$^+_1$ and 1$^-_2$, makes the angular distribution asymmetric.
  The present model gives the $E$2/$E$1 ratio $\sigma_{E2}/\sigma_{E1}= 9.0$, 2.3, 0.42, and 0.03 at $E_\gamma=8.0$, 8.5, 9.0, and 9.5 MeV.

\section{Summary}
  In this contribution, I have discussed the reaction mechanism of  $^{12}$C($\alpha$,$\gamma$)$^{16}$O, and have shown the theoretical result of the photodisintegration of $^{16}$O.

  The direct capture component of PM seems to describe the fundamental process of the stellar $^{12}$C($\alpha$,$\gamma$)$^{16}$O reaction.
  This stems from the $\alpha$+$^{12}$C molecular states, 2$^+_1$ and 1$^-_2$.
  The low-energy cross section is dominated by $E$2 transition, i.e. the tail of the subthreshold 2$^+_1$ state.
  The $E$1 transition is not strongly enhanced by the subthreshold 1$^-_1$ state.
  The weak coupling between two 1$^-$ states can be expected.
  The additional resonant contributions are found to be negligible.
  The direct capture component has been found to be predominant below the barrier.

  The photoelectric cross section of $^{16}$O($\gamma$,$\alpha$)$^{12}$C is predicted to be dominated by $E$2 excitation in the vicinity of the $\alpha$-particle threshold.
  The $\alpha$-particle angular distribution for $E_\gamma= 8.0$ -- 9.5 MeV is made from the interference between two molecular states.
  The future experimental projects will give the $E$2/$E$1 ratio and the reaction rates of $^{12}$C($\alpha$,$\gamma$)$^{16}$O, more accurately \cite{MMM,Gai14}.

  The author is grateful to Professors M.~Arnould, A.~Jorissen, K.~Takahashi, and H.~Utsunomiya for their hospitality during his stay at Universit\'e Libre de Bruxelles, and thanks Professors Y.~Ohnita and Y.~Sakuragi for their hospitality during his stay at Osaka City University.

\end{document}